\documentclass[floatfix,amsmath,amsfonts,amssymb,aps,twocolumn,pra]{revtex4-1}
\usepackage[pdftex]{graphicx}

\usepackage[pdftex]{hyperref}

\newcommand{\ee}{\mathrm{e}}
\newcommand{\du}{\mathrm{d}}
\newcommand{\dc}{\mathcal{D}}
\newcommand{\br}{\mathbf{r}}

\newcommand{\bk}{\mathbf{k}}

\newcommand{\bp}{\mathbf{p}}

\newcommand{\ii}{\mathrm{i}}

\newcommand{\avg}[1]{\langle #1 \rangle}
\newcommand{\kF}{k_\mathrm{F}}
\newcommand{\EF}{E_\mathrm{F}}
\newcommand{\Eb}{E_\mathrm{b}}
\newcommand{\rc}{r_\mathrm{c}}
\newcommand{\lz}{l_\mathrm{z}}
\newcommand{\wz}{\omega_\mathrm{z}}
\newcommand{\Reff}{R_\mathrm{eff}}

\newcommand{\mytimes}{\times}

\DeclareMathOperator{\sgn}{sgn}
\DeclareMathOperator{\arcsinh}{arcsinh}

\newcommand{\figref}[1]{Fig.~\ref{#1}}
\newcommand{\appref}[1]{Appendix~\ref{#1}}
\newcommand{\reference}[1]{Ref.~\cite{#1}}
\newcommand{\refs}[1]{Refs.~\cite{#1}}
\newcommand{\secref}[1]{Section~\ref{#1}}
\newcommand{\eqnref}[1]{Equation~(\ref{#1})}

\begin{document}

\title{Effective-range dependence of two-dimensional Fermi gases}
\author{L.M.~Schonenberg}
\affiliation{Cavendish~Laboratory, J.J.~Thomson~Avenue, Cambridge, CB3~0HE, United Kingdom}
\author{P.C.~Verpoort}
\affiliation{Cavendish~Laboratory, J.J.~Thomson~Avenue, Cambridge, CB3~0HE, United Kingdom}
\author{G.J.~Conduit}
\affiliation{Cavendish~Laboratory, J.J.~Thomson~Avenue, Cambridge, CB3~0HE, United Kingdom}
\date{\today}

\begin{abstract}
The Feshbach resonance provides precise control over the scattering
length and effective range of interactions between ultracold atoms. We
propose the ultratransferable pseudopotential to model effective
interaction ranges $-1.5 \leq \kF^2 \Reff^2 \leq 0$, where $\Reff$ is the 
effective range and $\kF$ is the Fermi wave vector, describing narrow
to broad Feshbach resonances. We develop a mean-field treatment and
exploit the pseudopotential to perform a variational and diffusion
Monte Carlo study of the ground state of the two-dimensional Fermi
gas, reporting on the ground-state energy, contact, condensate
fraction, momentum distribution, and pair-correlation functions as a
function of the effective interaction range across the BEC-BCS
crossover. The limit $\kF^2 \Reff^2 \to -\infty$ is a gas of
bosons with zero binding energy, whereas $\ln(\kF a) \to -\infty$
corresponds to noninteracting bosons with infinite binding energy. 
\end{abstract}

\maketitle

\section{Introduction}

Many discoveries in modern condensed matter physics have emerged in
two-dimensional systems, such as the quantum Hall effect
\cite{Lin2009}, the BKT transition \cite{Hadzibabic2006}, and high
temperature superconductivity \cite{Lee2006}. Recent experimental
advances allow for the realization of a
two-dimensional ultracold atomic gas by means of an anisotropic
optical trap that confines one dimension more tightly than the other
two \cite{Martiyanov2010,Gorlitz2001}. In combination with the
Feshbach resonance \cite{Chin2010,Bloch2008} this provides a platform for the
controlled study of interactions in the two-dimensional Fermi gas,
which has attracted considerable interest both experimentally
\cite{Sommer2012,Gunter2005,Martikainen2005,Orel2011,Boettcher2016}
and theoretically
\cite{Petrov2003,Botelho2006,Zhang2008,Conduit2008,Ngampruetikorn2013,Drummond2011,Boettcher2016}
for contact interactions. Here we study the BEC-BCS crossover as a
function of the effective range of the interaction.

The scattering of two interacting atoms at
low energies is described by the s-wave scattering phase shift $\delta(k)$, which
up to second order in the wave vector $k$ is parameterized by
\cite{Verhaar1984,Adhikari1986a},
\begin{equation}
  \label{eq:phaseshift-2d}
  \cot[\delta(k)] = \frac{2}{\pi} \ln(k a) + \frac{k^2 \Reff^2}{4}.
\end{equation}
Here $a$ is the scattering length and $\Reff$ the effective range. The
noninteracting gas has infinite scattering length, $a=\infty$, and the
contact interaction used in earlier theoretical works
\cite{Bertaina2011,Shi2015,Galea2016} is recovered in the limit of
zero effective range, $\Reff^2 = 0$.  The effective range for a 2D
resonance is shown to be related to the 3D effective range by $\kF^2
\Reff^2 \simeq \kF^2 \lz \Reff^{\mathrm{3D}}$, with Fermi wave vector
$\kF$ and the harmonic oscillator length of the tight confinement
direction is $\lz$. $\Reff^{\mathrm{3D}} < 0$ for narrow Feshbach
resonances and $\Reff^{\mathrm{3D}} \approx 0$ for broad Feshbach
resonances so that in typical experiments exploiting the full gamut of
available resonances, $-100 \lesssim \kF^2 \Reff^2 \lesssim 0$
\cite{Boettcher2016,Wang2011,Chin2010}.

In this paper, we extend the analysis of
\refs{Randeria1989,Randeria1990} to derive a mean-field theory that is
quantitatively correct in the limit of large negative effective
interaction range $\kF^2\Reff^2\ll-1$. This is complemented by
variational Monte Carlo (VMC) and diffusion Monte Carlo (DMC)
simulations in the strongly interacting regime, for which we develop
an ultratransferable pseudopotential (UTP) following
\refs{VonKeyserlingk2013,Bugnion2014,Whitehead2016a,Schonenberg2016} that is calibrated
to deliver both the correct scattering phase shift and correct binding
energy for the two-body bound state for $-1.5 \leq \kF^2 \Reff^2 \leq
0$. Exploiting the UTP, we first revisit the case $\Reff^2=0$
\cite{Bertaina2011,Shi2015,Galea2016}, which we next use as a concrete
basis to analyze $\Reff^2<0$ by considering the ground-state energy,
condensate fraction, momentum distribution, and pair-correlation
functions.
 
In \secref{meanfield-2d} we study the two-body problem and use it as
a building block for our mean-field theory of the many-body
problem. In \secref{pseudopotentials-2d} we propose the UTP and
demonstrate that its scattering phase shift and bound-state energy are
more accurate than the conventional potential well. We study how the
BEC-BCS crossover evolves as a function of the effective interaction
range using DMC in \secref{qmc-2d}. Finally, we provide a
discussion of our main findings in \secref{discussion-2d}.

\section{Mean-field theory}
\label{meanfield-2d}

We develop a mean-field theory for the ground state to provide a solid 
foundation for our investigations. The theory 
becomes exact in the noninteracting limit, $\Reff^2 \to
-\infty$. For less negative values of $\Reff^2$ fluctuations around
the mean-field solution increase and the mean-field theory remains 
qualitatively correct, but serves as a concrete base to compare with
our quantum Monte Carlo study of the strongly interacting regime. Our
two-dimensional treatment is analogous to the three-dimensional case
discussed in \reference{Gurarie2007}.

We use the two-channel model introduced in \reference{Timmermans1999}
and employ atomic units ($\hbar = m = 1$) throughout this text,
\begin{align}
  \label{eq:two-channel-hamiltonian-2d}
  \hat{H}^\mathrm{2-ch} = &\sum_{\bk,\sigma} \frac{k^2}{2}
  c_{\bk \sigma}^\dagger c_{\bk \sigma} +
  \sum_{\bp} \bigg(\epsilon_0 + \frac{p^2}{4} \bigg)
  b_{\bp}^\dagger b_{\bp} \nonumber \\
  & + \sum_{\bk,\bp} \frac{\lambda}{\sqrt{A}} (
  b_{\bp} c_{\frac{\bp}{2}+\bk \uparrow}^\dagger
    c_{\frac{\bp}{2}-\bk \downarrow}^\dagger + \mathrm{h.c.}).
\end{align}
$c_{\bk \sigma}^\dagger$ creates and $c_{\bk \sigma}$ annihilates a
fermion with momentum $\bk$ and spin $\sigma$ respectively. Similarly,
$b_{\bp}^\dagger$ creates and $b_{\bp}$ annihilates a boson with
momentum $\bp$. $\epsilon_0$ is the bare detuning of the bosonic mode,
$\lambda$ is the coupling between the fermionic and bosonic channels,
and $A$ is the area.

We first solve exactly for the two-body sector in this theory, relating the
model parameters $\lambda$ and $\epsilon_0$ to the physical scattering
length $a$ and effective range $\Reff$. We then study the many-body
problem in mean-field approximation, computing the chemical potential,
BCS energy gap, and ground-state energy.

\subsection{Two-body problem}

To calibrate the model parameters $\lambda$ and $\epsilon_0$, we compute the 
two-body scattering 
amplitude using the two-channel model Hamiltonian, and match the
result to the scattering amplitude corresponding to the desired
scattering phase shift.  We consider the scattering of a spin-up
fermion with momentum $\bp/2 + \bk$ and a spin-down fermion with
momentum $\bp/2 - \bk$, so that the center of mass momentum is $\bp$,
while the momentum in the center of mass frame is $\bk$.  The
scattering amplitude $f$ equals to the $T$-matrix \cite{Adhikari1986},
which is computed as the renormalized four-point vertex, $T(\bk,\bk')
= \Gamma(\bp/2+\bk,\bp/2-\bk,\bp/2+\bk',\bp/2-\bk')$.  As the contact
interaction has no angular dependence, we expect only s-wave
scattering so that the $T$-matrix depends only on the magnitude of the
relative momentum $k = |\bk| = |\bk'|$.

The T-matrix is computed as the sum of the geometric perturbation
series in $\lambda$,
\begin{equation}
  \label{eq:t-matrix-2d}
  T(k) = [(\lambda^2 D_0)^{-1} - \Pi]^{-1}.
\end{equation}
$D_0$ is the bosonic propagator evaluated at momentum $\bp$ and energy
$k^2+p^2/4$, and $\Pi$ is the polarization operator for
fermions with relative energy $k^2$,
\begin{align}
  \label{eq:shorthands-2d}
  D_0^{-1} =& k^2 - \epsilon_0 + \ii 0^+, \\
  \Pi =& -\frac{1}{4 \pi} \ln \bigg(1-\frac{2 \Lambda^2}{k^2} \bigg).
\end{align}
$0^+$ is an infinitesimal positive number and $\Lambda$ is a momentum
cutoff, required to regularize the integral over the relative momentum
of the two particles that diverges as the result of the contact
interaction between the fermionic and bosonic channels. Physically,
the regularization leads to a renormalization of the bare detuning
$\epsilon_0$ to give a physical detuning $\omega_0$,
\begin{equation}
  \label{eq:detuning-2d}
  \omega_0 = \epsilon_0-\frac{\lambda^2}{2 \pi} \ln(\Lambda/q^*),
\end{equation}
where $q^*$ is an arbitrary momentum scale that can be chosen at
convenience. 

Assuming the momentum cutoff to be arbitrarily
large, $\Lambda \gg k$, the scattering amplitude reads,
\begin{equation}
  \label{eq:scattering-amplitude-2d-2}
  f(k) = \frac{4}{-\frac{2}{\pi} \ln(\frac{k}{q^*}
    \ee^{2\pi\omega_0/ \lambda^2})+\frac{4k^2}{\lambda^2}+\ii}.
\end{equation}
The scattering amplitude is related to the phase shift as $f(k) =
4/\{\cot[\delta(k)] - \ii \}$ \cite{Adhikari1986}, so in terms of
the scattering length $a$ and effective range $\Reff$
\begin{equation}
  \label{eq:scattering-amplitude-2d}
  f(k) = \frac{4}{-\frac{2}{\pi} \ln(k a) - \frac{k^2
      \Reff^2}{4} + \ii}.
\end{equation}
Matching both expressions, we can express $\omega_0$ and $\lambda$ in
terms of $a$ and $\Reff$,
\begin{subequations}
\begin{align}
  \label{eq:detuning-scattering-length-2d}
  \omega_0 =& -\frac{8}{\pi \Reff^2} \ln(q^* a), \\
  \label{eq:g-effective-range-2d}
  \lambda^2 =& - \bigg(\frac{4}{\Reff}\bigg)^2.
\end{align}
\end{subequations}

The 2D scattering length $a$ and effective range $\Reff$ are related
to their 3D counterparts $a^\mathrm{3D}$ and $\Reff^\mathrm{3D}$
through the physical detuning, which is independent of dimensionality
and related to the experimental magnetic field. Equating the 3D detuning
$2/(\Reff^\mathrm{3D} a^\mathrm{3D})$ \cite{Gurarie2007} to the 2D detuning,
\eqref{eq:detuning-scattering-length-2d},
\begin{equation}
  \label{eq:scattering-lengths-2d-3d-b}
  a = (q^*)^{-1} \exp[-4 \Reff^2/(\pi a^\mathrm{3D} \Reff^\mathrm{3D})].
\end{equation}
This expression is of the same form as the one found by
\citet{Petrov2001} for particles confined to a two-dimensional plane
by a harmonic potential, $a \simeq 1.86 \, \lz \exp(-\sqrt{\pi/2}
\lz/a^\mathrm{3D})$, where the harmonic oscillator length in the
direction normal to the plane $\lz = 1/\sqrt{\wz}$ with $\wz$ the
oscillator frequency. Comparing both expressions, we find for the 2D
effective range of particles confined in a harmonic potential
\begin{equation}
  \label{eq:effective-range-2d-3d}
  \Reff^2 \simeq 0.984 \, \lz \Reff^\mathrm{3D}.
\end{equation}
Since $\Reff^\mathrm{3D} < 0$ in experiments, the quantity $\Reff^2$ is also 
negative.

The energy of the two-body bound state of a pair of fermions, i.e.,
the renormalized boson, $\Eb =k^2$ can be computed from the
corresponding pole in the scattering amplitude as
\begin{equation}
  \label{eq:bound-state-energy-2d}
  \Eb = \frac{4}{\pi \Reff^2} W_0 \bigg(- \frac{\pi \Reff^2}{4 a^2}\bigg)
\end{equation}
where $W_0$ is the principle branch of the Lambert-W function, defined
as the solutions to the equation $z = W(z \ee^z)$. In the limit
$\Reff^2 \to 0$ the equation reduces to $\Eb = -1/a^2$\footnote{Some
authors use an alternative definition of the scattering length, $a'=2a
\ee^{-\gamma}$ with $\gamma \approx 0.577$ Euler's constant. In this
case $\Eb=-\ee^{2\gamma}/4ma'^2$}, and in the limit $\Reff^2 \to
-\infty$, $\Eb = 0$. 

As is evident from \eqnref{eq:g-effective-range-2d}, $\Reff^2 \to
-\infty$ corresponds to the limit $\lambda \to 0$ where the fermions
do not interact with the bosons. In this case the two-body bound state
of a pair of fermions, with energy $\Eb=0$, is equal to a bare boson,
with energy equal to its bare detuning and indeed
$\epsilon_0 = \omega_0 = 0$. In the presence of interactions the
two-body bound state is a quasiparticle formed of a boson dressed by
fermionic fluctuations and its energy is therefore no longer equal to
the energy of a bare boson.

\subsection{Many-body theory}

Now that we have calibrated our model parameters to give the desired
two-body scattering properties, we turn to the many-particle theory
using the noninteracting limit $\lambda \to 0$ as a solid platform. 
The energy of a boson is zero in this limit, but the energy of
a fermion is positive because it has a finite kinetic energy due to
the Pauli exclusion principle. The ground state is therefore a BEC of
bosons that have no residual interactions with each other, which we
shall use as a concrete platform for the development of a perturbative
mean-field theory.

We consider the grand-canonical partition function expressed as a path
integral. After integrating out the quadratic fermion fields, the
partition function becomes $\mathcal{Z} = \int \dc \phi \dc
\bar{\phi} \, \exp(-S[\phi,\bar{\phi}])$, where the action,
\begin{widetext}
\begin{align}
  S[\phi,\bar{\phi}] = \int_0^\beta \du \tau \int \du^2 r
  \bar{\phi} \bigg(\partial_\tau + \epsilon_0 - 2 \mu -
    \frac{\nabla^2}{4m} \bigg) \phi - \ln \det \begin{pmatrix}
    \partial_\tau - \frac{\nabla^2}{2m} -\mu & \lambda \phi \\
    \lambda \bar{\phi} & \partial_\tau + \frac{\nabla^2}{2m} + \mu
    \end{pmatrix},
\end{align}
\end{widetext}
is a function of the bosonic field $\phi(\br,\tau)$. The imaginary
time integral runs up to the inverse temperature $\beta = 1/T$ and
$\mu$ is the chemical potential. Since two fermionic particles can
convert into a bosonic molecule through the interaction term
$\lambda$, $\mu$ couples to the total conserved particle density $n =
n_\mathrm{f} + 2 n_\mathrm{b}$ where $n_\mathrm{f}$ and $n_\mathrm{b}$
is the density of fermionic and bosonic particles respectively.

We use a mean-field approximation, replacing the path integral over
the bosonic field by a single real mean-field $\phi(\br,\tau)=B$ that
minimizes the action. At zero temperature the BCS-equation, obtained
from the condition $\delta S/ \delta \phi = 0$, and the number
equation, obtained as $n = -(T/A) \partial S[B]/\partial \mu$, where
the action is now a functional of the mean-field $B$, read,
\begin{subequations}
\begin{align}
  \label{eq:gap-number-equations-2d}
  \epsilon_0 -2\mu =& \frac{\lambda^2}{2} \int \frac{\du^2 k}{(2 \pi)^2}
                         \frac{1}{E(k)}, \\
  n =& 2 \bigg(\frac{\Delta}{\lambda}\bigg)^2 + \int \frac{\du^2 k}{(2 \pi)^2} \bigg(1 -
  \frac{\xi(k)}{E(k)} \bigg).
\end{align}
\end{subequations}
With $\Delta=\lambda B$ and the usual BCS expressions for $\xi(k)$ and $E(k)$,
\begin{subequations}
\begin{align}
  \xi(k) =& \frac{k^2}{2} - \mu, \\
  E(k) =& \sqrt{\xi^2(k) + \Delta^2}. 
\end{align}
\end{subequations}
The momentum integral in the gap equation diverges and is regularized
as done before by introducing the momentum cutoff $\Lambda$ and eliminating
the bare detuning $\epsilon_0$ in favor of the physical detuning
$\omega_0$
\begin{equation}
  \label{eq:gap-regularized-2d}
   \omega_0 - 2 \mu = \frac{\lambda^2}{2} \int \frac{\du^2 k}{(2 \pi)^2}
    \bigg[\frac{1}{E(k)} -\frac{2}{k^2}
    \Theta \bigg(\frac{k^2}{\kF^2}-1 \bigg) \bigg],
\end{equation}
where we set $q^*=\kF$, the Fermi momentum. 

The integrals can be performed analytically in terms of the
dimensionless ratio $\mu/\Delta$ to obtain our result for
$\Delta$ and $\mu$,
\begin{subequations}
\begin{align}
  \label{eq:gap-number-equations-crossover-2d}
  \omega_0- 2\mu =& \frac{\lambda^2}{4\pi} \bigg[\arcsinh \bigg(\frac{\mu}{\Delta}\bigg)
 +  \ln \bigg(\frac{\kF^2}{\Delta} \bigg) \bigg], \\
  n =& 2 \bigg(\frac{\Delta}{\lambda}\bigg)^2 + 
  \frac{1}{2\pi} [\sqrt{\Delta^2+\mu^2} 
  +\mu ].
\end{align}
\end{subequations}
The ground-state grand canonical potential is then computed as
$\lim_{T \to 0} T S[B]/A$. Converting from energy per unit area to
energy per particle and adding back the chemical potential, the ground
state energy per particle is
\begin{widetext}
\begin{align}
  \label{eq:energy-density-2d}
  E = & \frac{2\pi \Delta^2}{\kF^2 \lambda^2} (\omega_0 - 2 \mu) - \frac{1}{2 \kF^2} 
 ( \mu^2 + \sqrt{\Delta^2+\mu^2} ) -
 \frac{\Delta^2}{2 \kF^2} \bigg[ \arcsinh\bigg(\frac{\mu}{\Delta} \bigg) + \frac{1}{2} + 
     \ln \bigg(\frac{\kF^2}{\Delta} \bigg) \bigg] + \mu.
\end{align}
\end{widetext}

The BCS equations can be solved analytically in the limit of small
$\Delta$, described in \appref{bec-bcs-limits-2d}, and at the BEC-BCS
crossover point $\mu=0$ \cite{Ngampruetikorn2013}, in which we are
interested here.  After setting $\mu=0$ and eliminating $\lambda$ in
favor of $\Reff$, the gap and number equations reduce to
\begin{subequations}
\begin{align}
  \omega_0 =& -\frac{4}{\pi \Reff^2} \ln \bigg(\frac{\kF^2}{\Delta} \bigg), 
  \label{eq:19a}\\
  \Delta =& \frac {2}{\pi \Reff^2} [1-\sqrt{1-\pi (\kF \Reff)^2}].
  \label{eq:19b}
\end{align}
\end{subequations}
This shows that $\Delta \to 0$ when approaching the noninteracting
limit $\Reff^2 \to -\infty$ as expected. Furthermore in this limit, 
the density of the bosons $n_b = B^2 \to n/2$, confirming that all
particles convert into composite bosons. Combining Equations \eqref{eq:19a} and
\eqref{eq:19b}
with \eqnref{eq:detuning-scattering-length-2d}, the scattering length
at the crossover point, $\mu = 0$, is related to the effective range as
\begin{equation}
  \label{eq:crossover-line-2d}
  a = \sqrt{\pi/2} |\Reff| [\sqrt{1-\pi(\kF \Reff)^2}-1]^{-1/2},
\end{equation}
showing that the scattering length increases as $\Reff^2$ is reduced
while keeping the chemical potential fixed. We will use
these results to compare the mean-field prediction with our diffusion
Monte Carlo estimate for the ground-state energy at the BEC-BCS
crossover as a function of the effective range in \secref{qmc-2d}.

\section{Pseudopotentials}
\label{pseudopotentials-2d}

To address the full gamut of effective ranges, we turn to numerical Quantum 
Monte Carlo simulations. 
For our quantum Monte Carlo simulation of the strongly interacting
regime, we eliminate the need to simulate the bosonic particles by using a
single-channel Hamiltonian that only includes the spin $1/2$ fermions,
\begin{equation}
 \label{eq:hamiltonian-2d} 
\hat{H}^\mathrm{1-ch} = -\frac{1}{2} \sum_{i=1}^N \nabla_i^2 +
\sum_{i < j}^{N} V(r_{ij}).
\end{equation} 
$\nabla_i^2$ is the Laplacian with respect to the coordinates of
particle $i$, $N$ is the total number of particles, and we study equal
numbers of up and down spin particles. $r_{ij}$ is the distance
between particles $i$ and $j$, and $V$ is an attractive interaction
potential that acts between particles with opposite spins. The aim of
this section is to develop a real-space form $V(\br)$ that scatters a
pair of fermions with the desired s-wave scattering phase shift
characterized by the scattering length $a$ and effective range
$\Reff$.

\begin{figure}
\includegraphics[width=\linewidth]{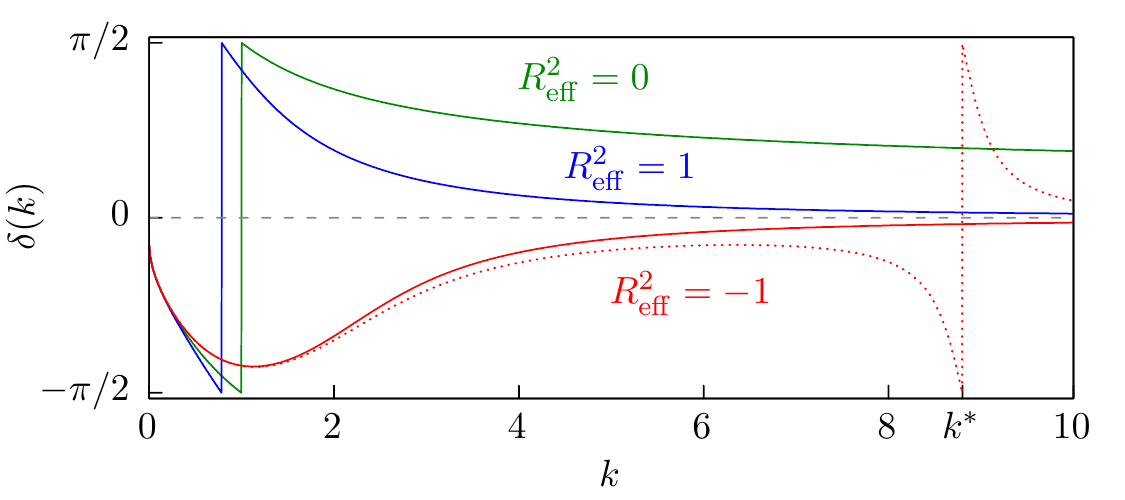}
\caption{(Color online) Scattering phase shift $\delta(k)$ for $a=1$
and $\Reff^2=\{-1,0,1\}$ (all lengths are in units of inverse
momentum). For $\Reff^2=-1$ the phase shift of a realistic potential
with the same low-energy scattering properties is indicated by the red
dotted line. The noninteracting phase shift is shown by the gray
dashed line.}
\label{fig:scattering-2d}
\end{figure}

The scattering phase shift for scattering length $a=1$ and effective range
squared $\Reff^2=\{-1,0,1\}$ is plotted in \figref{fig:scattering-2d}
(all length scales are in units of inverse momentum). The cases for
$\Reff^2 = 0$ and $\Reff^2=1$ are qualitatively similar, while the
case for $\Reff^2 = -1$ differs by the absence of the phase winding by
$\pi$. Furthermore, as $|\Reff^2|$ becomes larger, the phase shift
decays more rapidly towards zero for large $k$, as expected for the
noninteracting limit $|\Reff^2| \to \infty$.

The absence of the phase winding of $\pi$ for the case $\Reff^2 =-1$
has an important consequence because the number of phase windings is
related to the number of bound sates $n$ by Levinson's theorem \cite{Bolle1984},
\begin{equation}
\delta(0) - \delta(\infty)=n \pi.
\end{equation}
Attractive interactions in 2D \footnote{For the existence of the bound
  the mean-field definition of attractive, i.e.,  $\int V(\br) \du
  \br < 0$, is sufficient, which automatically includes all potentials
  satisfying the more stringent condition of uniform attractiveness, $V(\br)<0$.}
exhibit at least a single bound state \cite{Kocher1977,Coutinho1983},
so $n \geq 1$. The scattering phase shift of the desired attractive
interaction therefore should include at least a single phase winding,
but as it stands this is not the case for $\Reff^2 = -1$ because
$\delta(0) - \delta(\infty)=0$. A physical potential must however have a phase 
winding at $k^*$ as indicated by the dashed line in the figure, which 
corresponds to adding a higher order term to the expansion of 
$\cot(\delta(k))$. Similar to the 3D case reported in
\reference{Schonenberg2016}, this additional phase winding does not affect the 
phase shift at low momenta, and provided $k^*$ is much larger than any other
momentum scale in the system, i.e., the Fermi momentum $\kF$ for a
fermionic many-body system, does not alter the physics of the system
as the interacting particles cannot probe these high momentum
features. For $\Reff^2 > (4 a^2)/(\pi
\ee)$, Levinson's theorem has another important consequence as the
bound state energy, \eqnref{eq:bound-state-energy-2d}, does not exist
because $W_0$ does not exist for $z<-1/\ee$. The absence of the bound
state violates Levinson's theorem, which implies that no real space
potential exists in this regime.

To describe interactions with effective range $-1.5 \leq \kF^2
\Reff^2 \leq 0$, we develop a pseudopotential that is smooth and
extended in space to aid the numerical convergence, and accurately
reproduces the scattering phase shift and bound state energy. We first
discuss the potential well as it is commonly used to simulate Fermi
gases with contact interactions, before introducing the UTP
\cite{Bugnion2014,Lloyd-Williams2015,Whitehead2016,Whitehead2016a,Schonenberg2016}
as a model potential for both zero and finite effective interaction
range. Because of its high accuracy, wide spatial
extent, and smoothness, we select the UTP for our numerical study.

\subsection{Potential well}

A potential well was used in
\refs{Astrakharchik2004,Astrakharchik2005,Bertaina2011} to
model the contact interaction obtained in the zero effective-range limit $\Reff^2=0$,
\begin{align}
V(r)=\begin{cases}
-U,& r\le \rc, \\
0,& r>\rc,
\end{cases}
\end{align}
with depth $U$ and radius $\rc$. The depth $U$ can be tuned to give
the correct scattering length $a$, while the effective range $\Reff^2$
is proportional to $\rc^2$ and thus positive \cite{Adhikari1986}.  To
ensure that the effective range term is small, \citet{Bertaina2011}
used $\kF \rc = 2.5 \mytimes 10^{-3}$. The discontinuity of the
potential well at $\rc$ can be avoided by using a smooth form $V(r) =
a/\cosh^2(br)$ with $a<0$ \cite{Forbes2012,Galea2016}, but this does not change
the essence of the problem as the potential remains uniformly
attractive and must be deep and narrow to
ensure small $\Reff^2$. With a small effective radius both potentials
are difficult to handle numerically so we propose the UTP as an
alternative that allows $\Reff^2$ to be varied independently of
$\rc^2$.

\subsection{UTP}
\label{UTP-2d}

We now propose a pseudopotential that gives the precise scattering
phase shift and bound-state energy for $-1.5 \leq \kF^2 \Reff^2 \leq
0$. Furthermore, the potential is smooth and extended in space, easing the
application of numerical methods. Following
\refs{Bugnion2014,Whitehead2016a,Schonenberg2016}, we propose a UTP
that takes a polynomial form within a cutoff radius $\rc$,
\begin{align}
V^\mathrm{UTP\!}(r)\!=\!\begin{cases}
\!\left(\!1\!-\!\frac{r}{\rc}\!\right)^2 \!\left[ u_1\!\left( 1+\frac{2r}{\rc} 
\right)\! +\! \displaystyle\sum_{i=2}^{N_\mathrm{u}} u_i 
\left(\!\frac{r}{\rc}\!\right)^{\!i} \right]\!,&\!\!\!\!r\le\rc,\\
0,&\!\!\!\!r>\rc,
\end{cases}
\end{align}
where the $u_i$ are the $N_\mathrm{u}=3$ optimizable coefficients. The
term $(1-r/\rc)^2$ ensures that the UTP goes smoothly to
zero at $r=\rc$, and the component $u_1(1+2r/\rc)$ constrains the
pseudopotential to have zero gradient at particle coalescence
to ensure that the wave function is smooth.

The coefficients $\{u_i\}$ are optimized by solving the Schr\"odinger
equation for the two-body problem numerically
\cite{Whitehead2016a,Schonenberg2016}. We minimize a cost function
$F$ containing two terms: 1) The difference of
the logarithmic derivative of the pseudopotential wave function with
the exact wave function evaluated at the exact bound-state energy
$\Eb$ and the cutoff radius $\rc$. This term serves to obtain the
correct bound state-wave function, and therefore binding energy. 2) The
difference in scattering phase shift between the pseudopotential and
the exact expression, summed over angular momentum channels $l$ and
averaged over the Fermi sea $0 \leq k \leq \kF$ weighted by the density of 
scattering states in the center of mass frame $g(x)=8x\{1-\frac{2}{\pi}[x
\sqrt{1-x^2}+\arcsin(x)]\}$ \cite{Whitehead2016a},
\begin{align}
\label{eq:rms-phase-shift-2d}
F =& \rc^2 \bigg|\frac{\du [\ln(\psi^{\mathrm{UTP}})]}{\du
     r} - \frac{\du[\ln(\psi)]}{\du r} \bigg|_{E=\Eb,r=\rc}^2 \\ \nonumber
&+ \frac{1}{\pi} \sum_l \int_0^{\kF}
  \left| \delta_l^\mathrm{UTP}(k) - \delta_l(k) \right|^2 
g(k/\kF) \du k.
\end{align}
The prefactors $\rc^2$ and $1/\pi$ serve to make both terms
dimensionless. The upper momentum cutoff for the integral determines up to 
which momentum scale the UTP will accurately reflect the desired phase shift 
and its value influences features of the phase shift at high momenta including 
the value of $k^*$. We have confirmed that our results are insensitive to the 
value of the cutoff and $k^*$.

While increasing the cutoff radius $\rc$ improves numerical performance, it 
also introduces higher systematic error. In particular, the cutoff radius 
should be chosen less than the interparticle spacing so that three-body 
scattering events are rare; we therefore follow the approach by 
\refs{Bugnion2014,Whitehead2016a,Schonenberg2016} and set $\rc = 1/\kF$ to 
balance statistical and systematic errors in the QMC results.

\subsection{Comparison of potentials}
\label{comparison-2d}

\begin{figure}
\includegraphics[width=\linewidth]{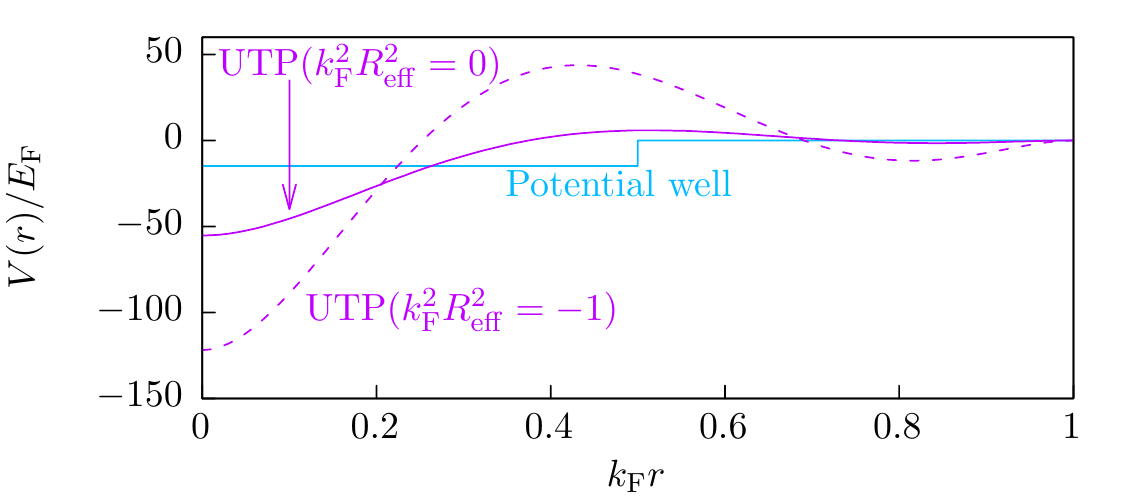}
\caption{(Color online) Plot of the potential well (blue), the UTP for
the contact interaction $(\kF \Reff)^2=0$ (solid purple), and the UTP
for finite effective range $(\kF \Reff)^2=-1$ (dashed purple),
normalized by the reciprocal Fermi energy $\EF$ as a function of the
dimensionless radius. All potentials are calibrated for interaction
strength $\ln(\kF a)=0$.}
\label{fig:potentials-2d}
\end{figure}

We compare the real-space forms of the potential well and UTP in
\figref{fig:potentials-2d} with $\ln(\kF a)=0$. We have chosen the
cutoff radius of the potential well $\rc = 1/(2\kF)$, such that its
spatial extent is similar to that of the UTP and so the computational
efficiency should be comparable. The potential well was used in
previous works \cite{Bertaina2011} to represent the contact
interaction, although it has $\Reff^2>0$. The UTP is shown for both the zero
range limit $\kF^2 \Reff^2 =0$ and for negative effective range
squared $\kF^2 \Reff^2 =-1$. Comparing the potential well with the UTP
for $\kF^2 \Reff^2 =0$, the potential well is shallower than the UTP
at small radius, but deeper at intermediate radius and furthermore
displays a discontinuity at the cutoff radius. In contrast, the UTP is
smooth throughout, easing the numerical optimization process of the
variational wave function. Reducing $\kF^2 \Reff^2$ from 0 to -1, the
UTP develops a potential barrier at intermediate radius.  The barrier
suppresses quantum tunneling between the composite two-fermion bound
state at small radius and the continuum of scattering states at large
radius. This physics was seen in the two channel model as $\lambda \rightarrow 
0$, and furthermore is reminiscent of the physics for negative effective range 
in three dimensions \cite{Schonenberg2016}.

\begin{figure}
\includegraphics[width=\linewidth]{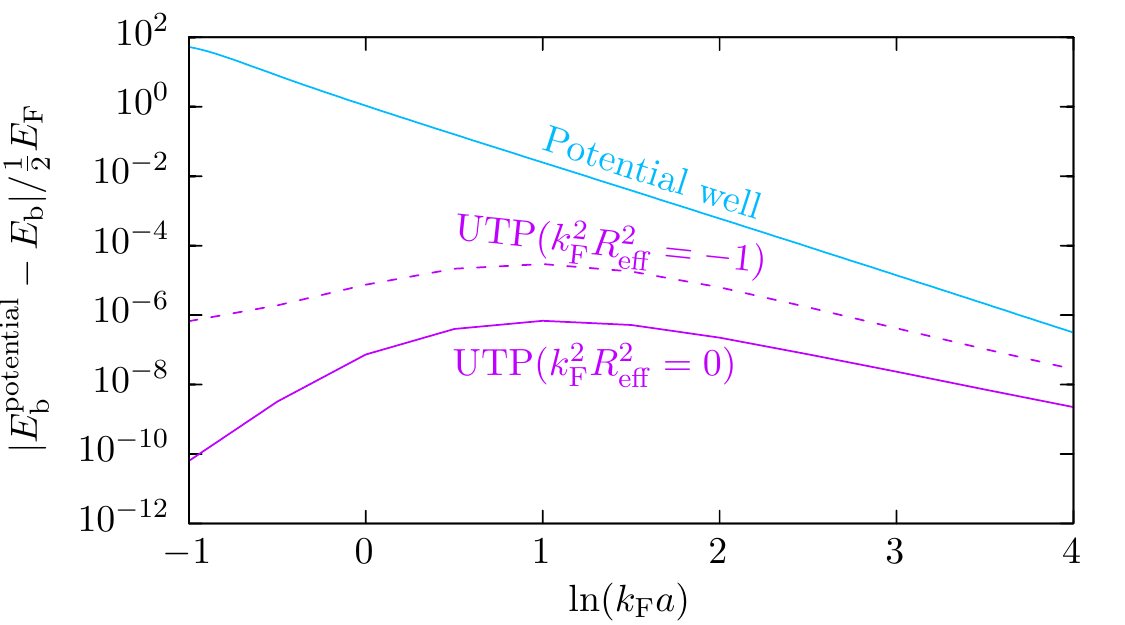}
\includegraphics[width=\linewidth]{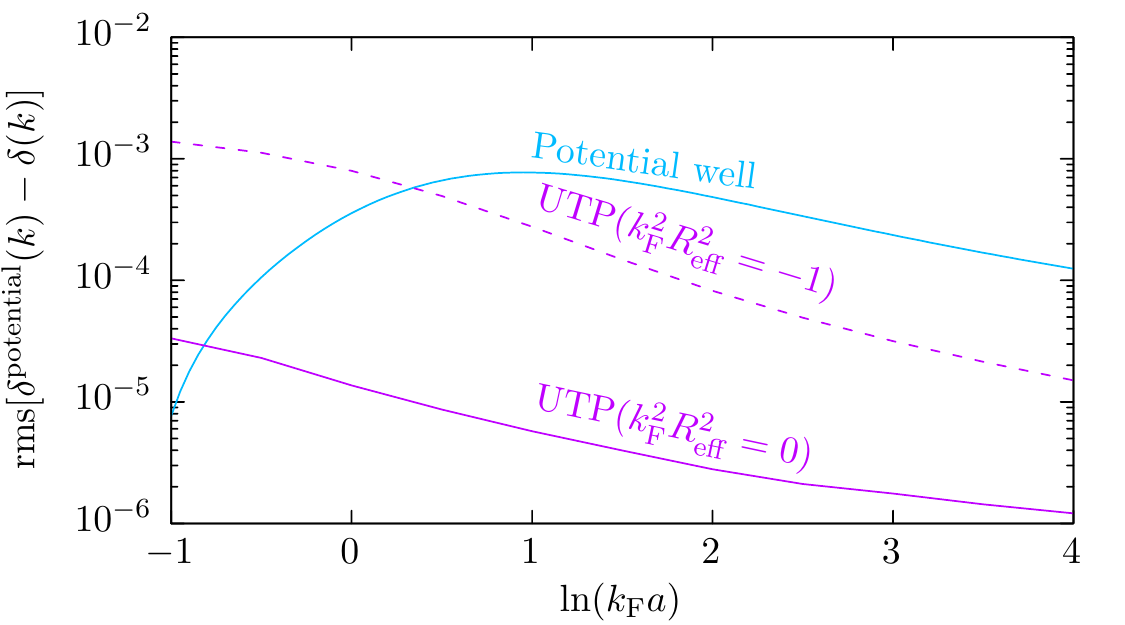}
\caption{(Color online) (Top) Difference in absolute value of the
bound-state energy of the potential and the exact bound-state energy
of the potential well (blue), the UTP for the contact interaction
$(\kF \Reff)^2=0$ (solid purple), and the UTP for finite effective
range $(\kF \Reff)^2=-1$ (dashed purple). (Bottom) Root mean square
(rms) scattering phase-shift error of the same potentials, with $k$
averaged over the interval from $0$ to the Fermi momentum $\kF$.}
\label{fig:phase-shift-2d}
\end{figure}

We next compare the accuracy of the bound-state energy and scattering
phase shift of the pseudopotentials, shown in
\figref{fig:phase-shift-2d}. The error in the bound-state energy for the
UTP is less than $10^{-6}$ in the zero effective-range limit and less then
$10^{-4}$ for $\kF^2 \Reff^2=-1$ for all interaction parameters
$\ln(\kF a)$. Moreover, the error in the bound-state energy decreases when
reducing the interaction parameter $\ln(\kF a)$ below zero. 
This regime corresponds to a BEC state of tightly-bound
bosons for the many-body system, and we therefore expect the
potentials to accurately describe this region. In contrast, the
bound-state energy error for the potential well is much larger, and
increases when approaching the BEC regime. Turning to the root mean
square (rms) scattering phase-shift error, accuracy is most important for the BCS
regime $\ln(\kF a) > 0$ of weakly bound particles, where scattering is
abundant. In this regime, we observe that the rms error is two orders
of magnitude smaller for the UTP in the zero effective-range limit compared with
the potential well, despite the fact that the potential well is
calibrated to yield the correct scattering length. The error of the
UTP for $\kF^2\Reff^2=-1$ is about two orders of magnitude larger at
$\ln(\kF a)=-1$ compared to the UTP for $\kF^2\Reff^2=0$, but
decreases to only one order of magnitude at $\ln(\kF a) =4$ as we move
towards the BCS regime.

We conclude that the UTP in the zero-range limit has a smaller error
in the bound-state energy and average scattering phase
shift than an analogous potential well.
Furthermore, the UTP evolves smoothly as a function of the effective
range, while the potential well cannot deliver negative $\kF^2 \Reff^2$.
We therefore select the UTP for our numerical study. 

\section{Quantum Monte Carlo}
\label{qmc-2d}

To calculate the ground-state properties of the Fermi gas in the
strongly interacting regime we use the
\textsc{casino} implementation of the fixed-node diffusion Monte Carlo
(DMC) algorithm \cite{Needs2010}. DMC is a Green's function projector method that
produces a variational upper bound on the ground-state energy,
depending only on the nodes of the trial wave function
\cite{Ceperley1980,Umrigar1993,Foulkes2001}. We start from the
Slater--Jastrow trial wave function $\Psi = \ee^J D$ introduced in
\reference{Schonenberg2016}. $D$ is a Slater determinant of $N/2$
pairing orbitals $\phi(\br_{ij})$, each holding an up- and down-spin
particle and $\br_{ij}$ the separation between them, and $\ee^J$ a Jastrow
factor that captures correlations between particles. The pairing
orbitals are formed of a linear combination of plane waves, compatible
with the nearly free electron gas in the BCS limit, and a polynomial
term, suitable for describing the weakly interacting composite bosons in
the BEC regime.  We use a backflow transformation to capture many-body
correlations in the pairing orbitals \cite{LopezRios2006}.  The trial
wave function includes a total of 33-39 parameters depending on the
number of particles simulated, which we optimize first using
variational Monte Carlo (VMC) before using it as input for our DMC
calculations.

We calculate the ground-state wave function for systems with 26 and 58
particles as in \refs{Bertaina2011,Galea2016}, and also for a system
of 98 particles to allow us to accurately extrapolate to the
thermodynamic limit. We also extrapolate to zero time-step and infinite
walker populations; details are provided in
\appref{dmc-extrapolations-2d}. We expect that the use of a quadratic DMC
algorithm would give similar results
\cite{Mella2000,Sarsa2002}. Expectation values of operators that do
not commute with the Hamiltonian are computed using the extrapolated
estimator $\avg{\hat{A}} = 2 \avg{\hat{A}}_\mathrm{DMC} -
\avg{\hat{A}}_\mathrm{VMC}$, such that the residual bias is quadratic in the
difference between the VMC and DMC wave functions for the part of the operator 
that is local in position space \footnote{Note that although Ceperley and Kalos 
\cite{Ceperley1986} report that the extrapolated estimate works in practice for 
matrix elements off-diagonal in position space, such as the momentum 
distribution and condensate fraction, their formal derivation breaks down for 
nonlocal operators.} \cite{Ceperley1986,Nightingale1999}. The extrapolated 
estimates are within the statistical error bars of the bare DMC estimates 
unless indicated otherwise, and we expect residual errors to be small.

In the limit $\kF^2 \Reff^2 = 0$ the Slater-Jastrow trial wave
function captures at least 92\% of the correlation energy, defined as
the difference between the Hartree-Fock and DMC ground state energy,
which is raised to 96\% using backflow transformations. For finite
effective ranges backflow transformations are especially important, as
the amount of correlation energy captured at the BEC-BCS crossover
point without backflow reduces from 95\% to 91\% while decreasing
$\kF^2 \Reff^2$ from 0 to -1.5, but remains at a constant 97\% using
backflow transformations.

\subsection{Zero-range limit}
\label{energy-2d}

\begin{figure}
\includegraphics[width=\linewidth]{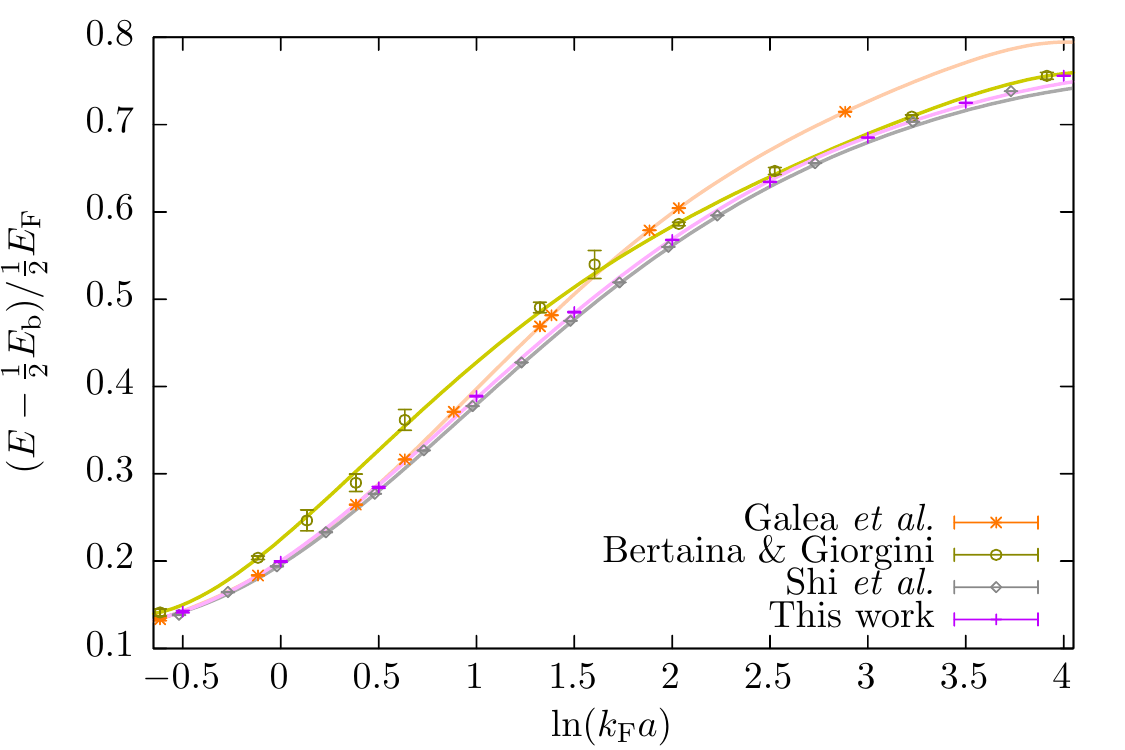}
\caption{(Color online) Ground-state energy per particle minus half
the two-body binding energy, divided by the energy per particle in a
noninteracting gas, as a function of the interaction parameter
$\ln(\kF a)$ across the BEC-BCS crossover. Next to our result we show
the results from \refs{Bertaina2011,Shi2015,Galea2016}.}
\label{fig:zero-range-2d}
\end{figure}

To demonstrate the accuracy of the proposed pseudopotential in
combination with our trial wave function, we first 
explore the ground-state energy of the gas across the BEC-BCS
crossover in the zero effective-range limit. This limit has been studied before
using DMC methods by \refs{Bertaina2011,Galea2016} and also using the
auxiliary-field Quantum Monte Carlo (QMC) method that is free from the
sign-problem for spin-balanced systems with attractive interactions by
\reference{Shi2015}. The BEC-BCS crossover is parameterized in 2D by
the interaction parameter $\ln(\kF a)$, which is inversely
proportional to the mean-field interaction strength 
\cite{Randeria1989,Randeria1990}. We study the
ground-state energy per particle $E$ minus half the two-body binding
energy $\Eb$, normalized by the energy per particle of a noninteracting gas
$\EF/2$. \figref{fig:zero-range-2d} shows that the relative energy increases
smoothly as the interaction parameter $\ln(\kF a)$ is increased from
negative values on the BEC side to positive values on the BCS
side. The polynomial fits to the data points are
obtained by explicitly taking into account the asymptotic functional
forms in the BEC and BCS limits as detailed in
\refs{Shi2015,Galea2016}.

By virtue of our pairing orbital that can smoothly interpolate
between the BCS and BEC limits we obtain a trial wave function that
provides the lowest DMC upper bound on the ground-state energy to
date. We benefit from our smooth pseudopotential in the regime $0
\lesssim \ln(\kF a) \lesssim 2$ where interactions are strong, while
for $2 \lesssim \ln(\kF a)$ we find that the finite size correction
leads to a significant reduction of the ground-state energy, and our
results are therefore lower than those reported by \citet{Galea2016}
(see \appref{finitesize-2d} for details). The reported DMC energies are
close to the auxiliary-field QMC results from \citet{Shi2015},
indicating that the fixed-node error is small.

\begin{figure}
\includegraphics[width=\linewidth]{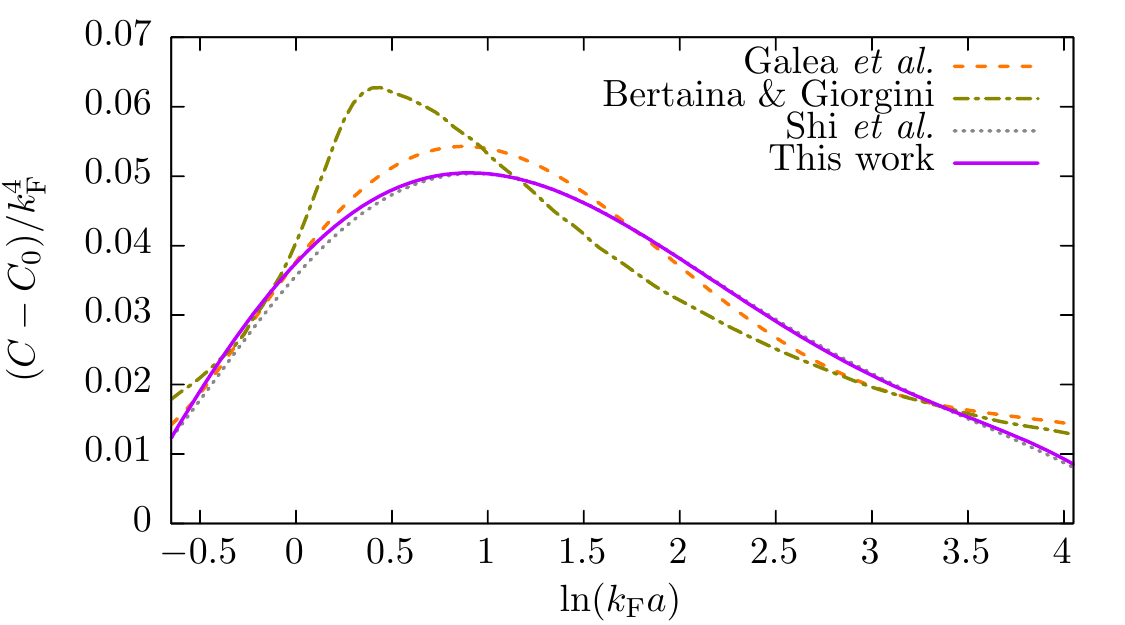}
\caption{(Color online) Contact minus the contact contribution from
the molecular bound state, normalized by the fourth power of the Fermi
wave vector. For comparison we also show the results from
\refs{Bertaina2011,Shi2015,Galea2016}.}
\label{fig:contact-zero-range-2d}
\end{figure}

An important ramification of the contact interaction in the
zero-range limit is the universal constant called the contact $C$,
which for example describes the high-momentum tail of the momentum
distribution, $n(k) \sim C/k^4$ \cite{Tan2008,*Tan2008a,*Tan2008b}. It
is proportional to the derivative of the equation of state,
\begin{equation}
  \label{eq:contact}
  \frac{C-C_0}{\kF^4} = \frac{1}{2} \frac{\du[E/\EF]}{\du
    [\ln(\kF a)]} -  \frac{1}{2} \frac{\du[(\Eb/2)/\EF]}{\du [\ln(\kF a)]}.
\end{equation}
Here $C_0$ is the contribution to the contact from the composite boson
that occurs at the mean-field level \cite{Bertaina2011}. The contact is shown in
\figref{fig:contact-zero-range-2d} and attains a maximum value 
at $\ln(\kF a) \approx 0.8$. As a result of our lower
upper bound on the ground state energy, our reported maximum value of
the contact is lower than earlier DMC studies
\cite{Bertaina2011,Galea2016} and agrees well with the auxiliary-field
QMC result from \citet{Shi2015}.

\subsubsection{Condensate fraction}

\begin{figure}
\includegraphics[width=\linewidth]{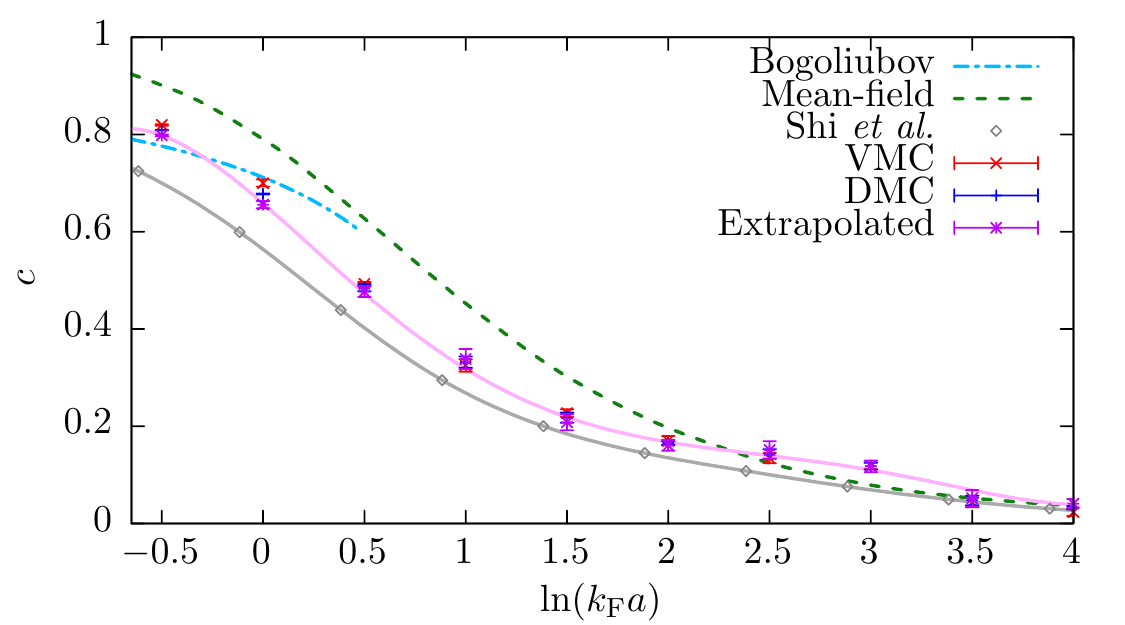}
\caption{(Color online) Condensate fraction as a function of the
interaction parameter $\ln(\kF a)$. We show the VMC, DMC and
extrapolated estimates, along with the auxiliary-field quantum Monte
Carlo result from \citet{Shi2015}. Also shown are the perturbative Bogoliubov
theory for the BEC limit \cite{Schick1971} and the mean-field prediction
\cite{Salasnich2007}.}
\label{fig:condfrac-zero-range-2d}
\end{figure}

A defining feature of a superconductor is the existence of a
condensate that correlates pairs of fermions with opposite spins
irrespective of the distance between them. Correlations between
particles are naturally captured in the two-body density matrix,
\begin{equation}
  \label{eq:2bdm} \rho_{\alpha \beta}^{(2)}(\br_1', \br_2';
\br_1, \br_2) = \langle c_\alpha^\dagger(\br_1')
c_\beta^\dagger(\br_2') c_\beta(\br_2)
c_\alpha(\br_1) \rangle,
\end{equation}
where $c_\alpha^{\dagger}(\br)$ is the fermionic creation and
$c_\alpha(\br)$ the annihilation operator for a particle with spin
$\alpha$ at position $\br$. The condensate fraction is defined as
$c=2n_0/n$, where the condensate density $n_0$ is the largest
eigenvalue of the two-body density matrix for particles with opposite
spins and $n$ is the density \cite{Leggett2006}, which we compute
using the improved estimator of \reference{Schonenberg2016}.

We show the extrapolated estimate together with the bare VMC and DMC
estimates in \figref{fig:condfrac-zero-range-2d}. For comparison we also plot
the Bogoliubov perturbation theory for the BEC limit \cite{Schick1971}
and the mean-field prediction \cite{Salasnich2007}. Furthermore, we show the 
auxiliary-field QMC results of \citet{Shi2015}. Starting from the BCS
regime where $\ln(\kF a)$ is large so 
interactions are weak, the condensate fraction increases exponentially
as $\ln(\kF a)$ is reduced. Our result is higher than the mean-field
prediction for $\ln(\kF a) \gtrsim 2.4$, while for $\ln(\kF a)
\lesssim 2.4$ the computed condensate fraction is lower than the
mean-field result and approaches the Bogoliubov result near $\ln(\kF
a) =-0.5$. The results computed using the extrapolated estimator are
similar to the bare VMC and DMC results, except near the BEC-BCS
crossover point $\ln(\kF a)=0$ where the correction is
significant. Our results are higher than those reported by
\citet{Shi2015}, although both results converge to the mean-field
theory as $\ln(\kF a) \to 4$. The disagreement could be the result of
the fixed-node approximation employed in this work, or the lattice
structure and corresponding breaking of rotational symmetry introduced
by \citet{Shi2015}.

\subsection{Finite effective range}

\begin{figure}
\includegraphics[width=\linewidth]{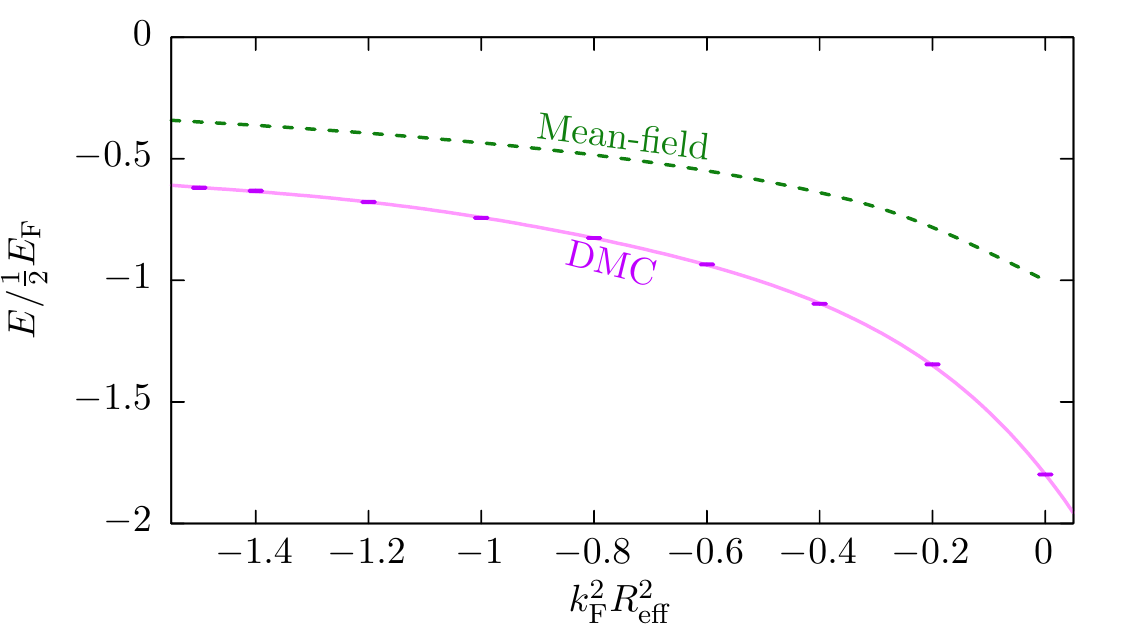}
\caption{(Color online) DMC ground-state energy per particle in units
of that of a noninteracting gas as a function of the dimensionless
effective range squared $\kF^2 \Reff^2$ shown by the solid purple
line. The mean-field result is shown by the dashed green line.}
\label{fig:energy-resqr-2d}
\end{figure}

Having demonstrated how the accurate scattering properties,
smoothness, and spatial extent of our pseudopotential result in
precise quantum Monte Carlo results in the familiar zero
effective-range limit, we now exploit the UTP to analyze interactions
with a finite effective range. We study the ground-state wave function
as a function of the effective range along the BEC-BCS crossover by
adjusting $a$ to follow the trajectory defined by $\mu=0$ in mean-field
approximation. We expect the actual trajectory of $\mu=0$ when calculated 
beyond mean-field approximation to track different values of $a$, as has been 
found for $\kF^2 \Reff^2=0$~\cite{Bertaina2011,Galea2016}. At finite effective 
range, we expect a similar shift, which approaches zero as 
$\kF^2 \Reff^2\rightarrow-\infty$, where the mean-field theory becomes exact.

The DMC ground-state energy is shown as a function of the effective
interaction-range squared $\kF^2 \Reff^2$ in
\figref{fig:energy-resqr-2d} together with the mean-field
prediction. Starting at the point $\kF^2 \Reff^2 = 0$ considered
before, we observe that the DMC estimate considerably improves the
mean-field estimate $E= -\EF/2$ by including additional correlations
between particles, reducing the energy to $E=-1.797(1) \EF/2$. As the
effective range becomes more negative, correlations beyond the
mean-field level diminish and the DMC result approaches the mean-field
result. The energy increases towards zero in the limit $\kF^2 \Reff^2
\to -\infty$ where the ground state is a condensate of noninteracting
composite bosons, each with zero internal binding energy.

\subsubsection{Condensate fraction}

\begin{figure}
\includegraphics[width=\linewidth]{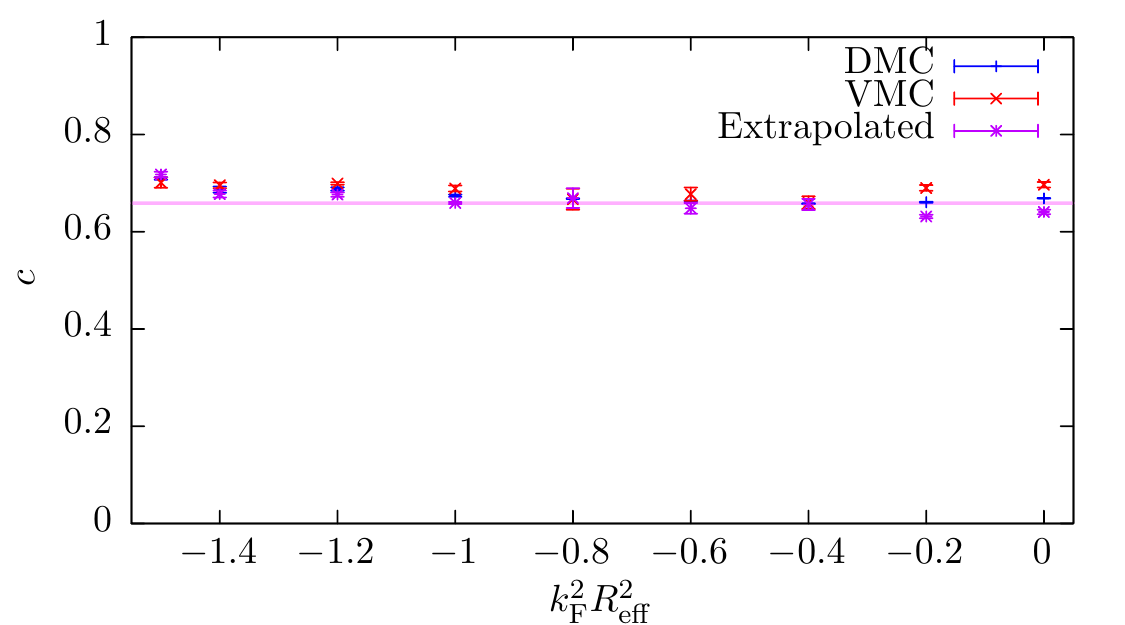}
\caption{(Color online) Condensate fraction as a function of the
  dimensionless effective
  range squared. We show the extrapolated result, as well as the bare
  VMC and DMC estimates.}
\label{fig:cond-frac-resqr-2d}
\end{figure}

The VMC, DMC, and extrapolated estimates for the condensate fraction
are shown as a function of $\kF^2 \Reff^2$ in
\figref{fig:cond-frac-resqr-2d}. The extrapolated estimator agrees
with the VMC and DMC results, except near the zero effective-range
limit considered before.  We observe that at the BEC-BCS crossover point in mean-field
approximation the condensate fraction remains constant at $c \approx
0.7$ for values of the effective range squared $-1.5 \leq \kF^2
\Reff^2 \leq 0$. This is consistent with our choice to tune $a$ and $\Reff$ to 
run along the BEC-BCS cross-over line, and is also observed in a three-dimensional 
system where a vanishing slope is seen at small negative effective range 
\cite{Schonenberg2016}. For large negative values of $\kF^2 \Reff^2$, we
expect the condensate fraction to decline as the mean-field theory
predicts $c \to 0$ in the limit $\kF^2 \Reff^2 \to -\infty$ where the 
theory becomes noninteracting and correlations vanish.

\subsubsection{Momentum distribution}

\begin{figure}
\includegraphics[width=\linewidth]{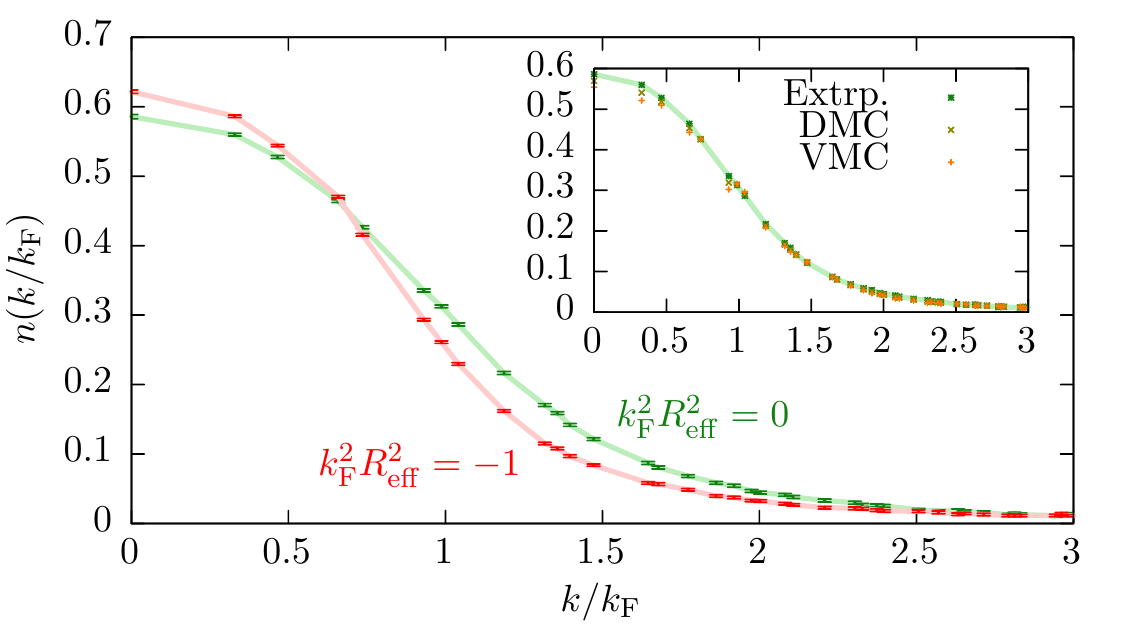}
\includegraphics[width=\linewidth]{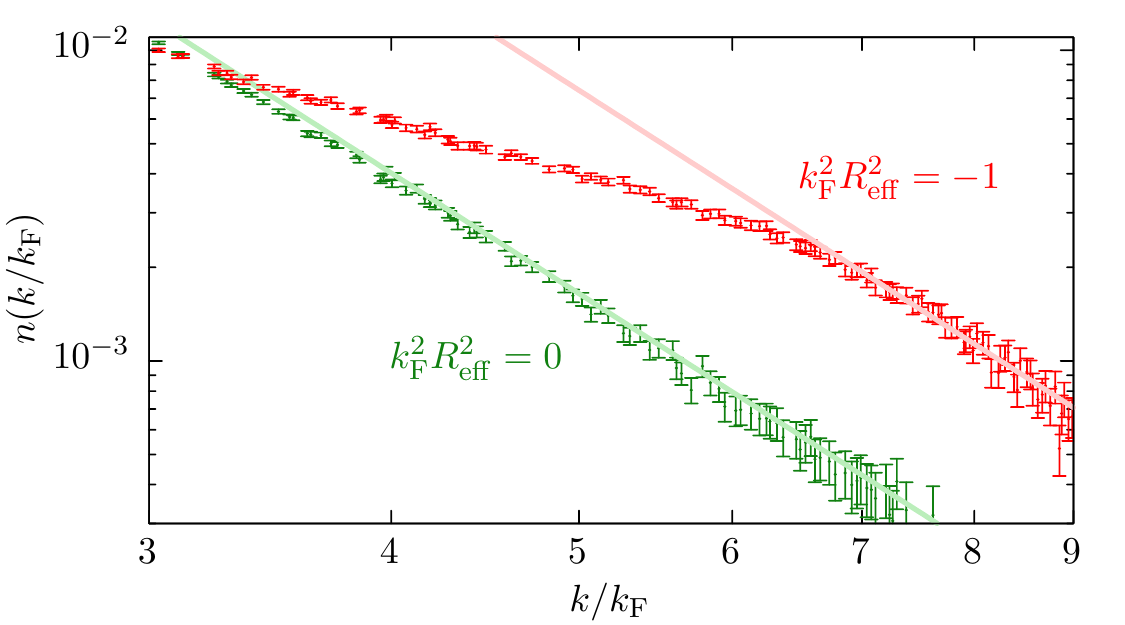}
\caption{(Color online) (Top) Extrapolated data for the momentum distribution 
$n(k)$ for $\kF^2 \Reff^2 \in \{0,-1\}$. The inset additionally shows DMC, VMC, 
and extrapolated results separately for $\kF^2 \Reff^2 = 0$. (Bottom) Tail of 
the momentum distribution on logarithmic axes, with the lines
indicating $n(k) \sim 1/k^4$.}
\label{fig:momdist-2d}
\end{figure}

\figref{fig:momdist-2d} shows the results of the extrapolated estimator 
for the momentum distribution $n(k)$ for $\kF^2\Reff^2 \in \{0,-1\}$ 
with an inset plot that shows DMC, VMC, and extrapolated data separately 
for $\kF^2\Reff^2 = 0$ (top), and the tail of the distributions for $k > 3.0 
\kF$ on logarithmic axes (bottom).

Both momentum distributions are different from a noninteracting Fermi
distribution, similar to what was found by \citet{Shi2015}
for $\kF^2 \Reff^2 =0$. The distribution has a wide-spread tail for
large momenta, and it is significantly reduced for momenta less than
$\kF$, which illustrates the formation of composite bosons. We expect
the high momentum components induced by the composite bosons to be accurately 
described by our UTP as it delivers the
correct two-body binding energy.

The extrapolation leads to small corrections only, with the most
significant corrections occurring near zero and the Fermi
momentum. This reflects the high quality of our trial wavefunction and
we expect residual errors to be small.

For $\kF^2 \Reff^2 =0$, a curve of $n(k) = C/k^4$ is shown in the logarithmic 
plot, where the value of $C/\kF^4=1.03$ has been extracted from the results 
that we obtained from Eq. \eqref{eq:contact}. The plotted line agrees well with
the QMC data within error bars and this is expected for $\kF^2\Reff^2 = 0$. For 
$\kF^2\Reff^2 = -1$, the tail of the momentum distribution decays less
rapidly before the $n \sim k^{-4}$ regime is entered at $k/\kF \approx
7$. A least-squares fit of $n(k) = C/k^4$ taking into account data for
$k/\kF > 7$ determines $C/\kF^4=4.64(4)$, and agrees well with the data
within error bars. The rise of the contact is expected as $\kF^2 \Reff^2$
is reduced due to the conversion of fermions to bosons,
akin to the 3D case \cite{Schonenberg2016}.  The $n(k) \sim
k^{-4}$ dependency for arbitrary effective range was demonstrated by
\refs{Braaten2008,Schonenberg2016} in 3D, and the QMC results studied
here suggest that this extends to 2D. In both cases, we expect the $n(k) \sim
k^{-4}$ dependency to hold only up to a certain momentum scale, and in 
particular not beyond the momentum scale up to which the UTP was optimized.

\subsubsection{Pair-correlation function}

\begin{figure}
\includegraphics[width=\linewidth]{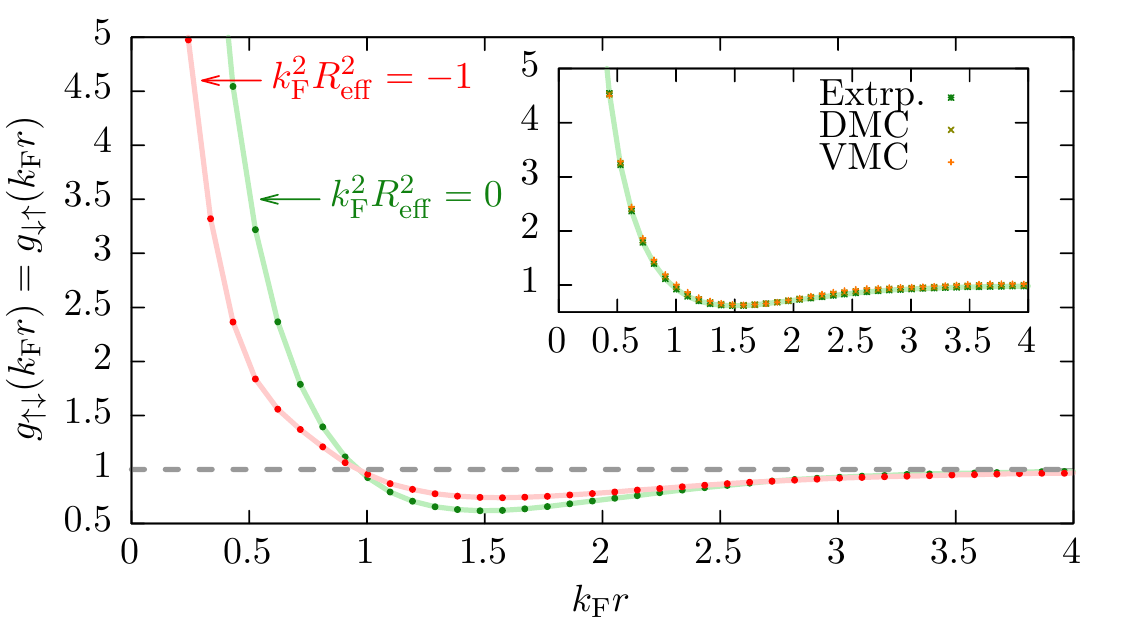}
\includegraphics[width=\linewidth]{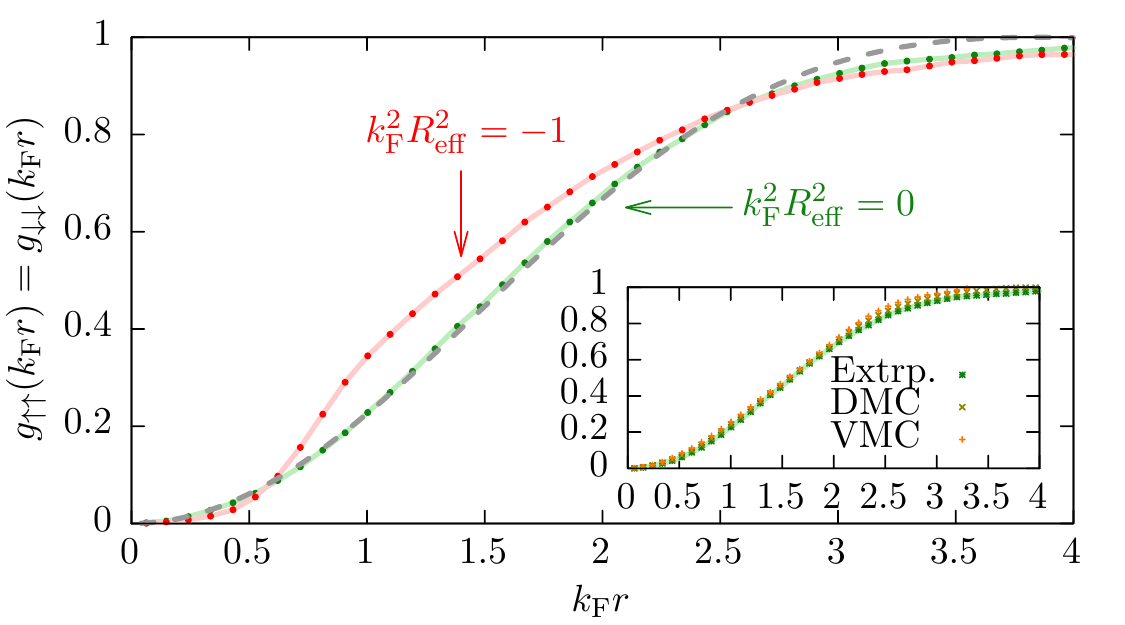}
\caption{(Color online) Pair-correlation function for opposite (top) and
equal (bottom) spins for $\kF^2 \Reff^2 \in \{-1,0\}$. The noninteracting
correlation function for fermions is indicated by the gray dashed line. The insets show 
DMC, VMC, and extrapolated data separately.}
\label{fig:paircorr-2d}
\end{figure}

The pair-correlation function (PCF) for two fermions is obtained from
the two-body density matrix as $g_{\alpha \beta}(\br) = \rho_{\alpha
\beta}^{(2)}(\br'+\br,\br';\br'+\br,\br')/(n/2)^2$.
\figref{fig:paircorr-2d} shows the results of the extrapolated estimator
for the PCF for opposite (top) and equal (bottom) spins for $\kF^2
\Reff^2 \in \{-1,0\}$ together with the result for noninteracting
fermions. The insets show DMC, VMC, and extrapolated data separately
for $\kF^2 \Reff^2 = 0$.

For opposite spins, we correct the PCF for short-range effects due to the 
particular form of the UTP \cite{Lloyd-Williams2015,Schonenberg2016}
\begin{equation}
  g_{\uparrow \downarrow}(r) = 
  \frac{g_{\uparrow\downarrow}^{\mathrm{2-body, exact}}(r)}{g_{\uparrow
      \downarrow}^{\mathrm{2-body, UTP}}(r)}
  g_{\uparrow \downarrow}^{\mathrm{QMC}}(r),
\end{equation}
where $g_{\uparrow\downarrow}^{\mathrm{2-body,\{exact,UTP\}}}(r)$ are the 
PCF for the two-body problem computed using the exact and UTP wave 
functions respectively, and $g_{\uparrow \downarrow}^{\mathrm{QMC}}(r)$ is the 
uncorrected QMC result for the PCF. Since our UTP is norm-conserving 
\cite{Bugnion2014,Whitehead2016a,Schonenberg2016}, no correction is necessary 
outside of the interaction region. Similar to the momentum distribution, 
the extrapolation from DMC and VMC data results in minor corrections only.

The PCF for opposite-spin species shows a significant increase at
short distances ($\kF r < 1$). This is due to the attractive
potential, and indicates the formation of composite bosons. The PCF is
reduced accordingly for intermediate distances ($\kF r \approx 1.5$). When
$\Reff^2$ becomes more negative, the fermions become more tightly
bound into bosons, but the residual interactions between two composite
bosons diminishes. As the conversion of fermions into noninteracting
point-like bosons progresses, the correlations between two fermions are
limited to a decreasing distance as is illustrated in the figure for
$\kF^2 \Reff^2 = -1$.

For equal spins, the result for $\kF^2 \Reff^2 = 0$ is similar to the
noninteracting case, which comprises the exchange-correlation hole due
to Pauli exclusion. When decreasing the effective range to $\kF^2
\Reff^2 = -1$, the fermions are more tightly bound as composite bosons
and the PCF is increased at short distances, reducing the distortion
caused by the exchange-correlation hole and approaching the PCF of
noninteracting bosons.

\section{Discussion}
\label{discussion-2d}

We investigated the many-body ground state of a two-dimensional Fermi
gas across the BEC-BCS crossover as a function of the effective
interaction range. A mean-field theory was developed showing that in
the limit $\kF \Reff^2 \to -\infty$ the fermions gain energy by
pairing into composite bosons in order to escape from the Fermi
sea. This should be contrasted with the noninteracting BEC found in
the limit $\ln(\kF a) \to -\infty$, where the binding energy of the
fermions diverges. Quantitatively accurate results are obtained for
the superconducting energy gap, chemical potential, and ground state
energy using the mean-field theory for $\kF^2 \Reff^2 \ll -1$.

To study the strongly interacting regime $-1.5 \leq \kF^2 \Reff^2 \leq
0$ using DMC, we proposed the ultratransferable
pseudopotential that produces the correct scattering phase shift and
bound state energy. We first revisited the case $\kF^2 \Reff^2=0$,
where we obtained the lowest DMC variational estimate of the
ground-state energy and confirmed that the fixed-node error is
small. We showed that the ground-state energy approaches the
mean-field prediction when the effective range is reduced, and showed
signatures for the formation of composite bosons in the momentum
distribution and pair-correlation functions. We confirmed the tail of
the momentum distribution $\sim 1/k^4$ for $\kF^2 \Reff^2 = 0$ and
demonstrated the same asymptotic form holds for $\kF^2 \Reff^2 < 0$,
as was previously observed in 3D \cite{Schonenberg2016,Braaten2008}.

While the mean-field theory provides qualitative insights into finite
interaction range effects and is also quantitatively trustworthy for
$\kF^2 \Reff^2 \ll -1$, we expect that our proposed ultratransferable
pseudopotential will be useful for future quantitative studies of the
strongly interacting regime for smaller negative values of
$\kF^2 \Reff^2$. The results presented for zero and finite effective ranges
should be relevant to ultracold-atom experiments using broad and narrow
Feshbach resonances respectively.

The software to generate the UTP and the data used in this work are
available online \cite{Schonenberg2016c}.

\acknowledgments {The authors thank Thomas Whitehead, Richard Needs, Stefano
Giorgini, Jordi Boronat, and Neil Drummond for useful discussions. The authors
acknowledge the financial support of the EPSRC Grant no. [EP/J017639/1], L.M.S.
acknowledges financial support from the Cambridge European Trust,
Cambridge Philosophical Society, VSB
Fonds, and the Prins Bernhard Cultuurfonds, and G.J.C. acknowledges the
financial support of the Royal Society and Gonville \& Caius College.
Computational facilities were provided by the University of Cambridge
High Performance Computing Service.}

\begin{appendix}

\section{BEC \& BCS limits at finite interaction range}
\label{bec-bcs-limits-2d}

In this section, we consider the limits of a small energy gap $\Delta$
corresponding to the BEC (negative chemical potential) and BCS
(positive chemical potential) regimes.

\paragraph{BEC-limit}
The chemical potential approaches $\Eb/2$ and is
therefore negative, so $\sgn(\mu/\Delta)=-1$, while its magnitude is large, so 
$|\mu/\Delta| \gg 1$. Expanding around $\mu/\Delta=-\infty$, the gap and number
equations read
\begin{subequations}
\begin{align}
  \label{eq:gap-number-bec-2d}
  B =& \frac{2|\mu|}{\lambda}
       \sqrt{-\frac{4\pi}{\lambda^2}(\omega_0-2\mu) +
       \ln\bigg(\frac{\kF^2}{2 |\mu|}\bigg)}, \\
  n =& \bigg(2 + \frac{\lambda^2}{4|\mu|} \bigg) B^2.
\end{align}
\end{subequations}

\paragraph{BCS-limit}
In the BCS-limit the chemical potential approximately equals the Fermi
energy $\mu \simeq \EF$, so $\sgn(\mu/\Delta)=1$ while
the gap is exponentially weak so $|\mu/\Delta| \gg 1$. 
Expanding the gap and number equations we find
\begin{subequations}
\begin{align}
  \label{eq:gap-number-bcs-2d}
  B =& \sqrt{2 \kF^2 \mu} \exp \bigg[-\frac{2\pi}
       {\lambda^2}(\omega_o-2 \mu) \bigg], \\
  n =& 2 B^2 + \frac{\mu}{\pi}.
\end{align}
\end{subequations}
Since $B$ is exponentially small, the number equation confirms that
the chemical potential approximately equals the Fermi energy as
$n=\kF^2/(2\pi)$.

\section{DMC extrapolations}
\label{dmc-extrapolations-2d}

To accurately extract the ground-state energy, it is important to
extrapolate to zero time step and infinite walker population, and to
the thermodynamic limit.

\subsection{Time step and walker population extrapolation}
\label{timestep-2d}

\begin{figure}
\includegraphics[width=\linewidth]{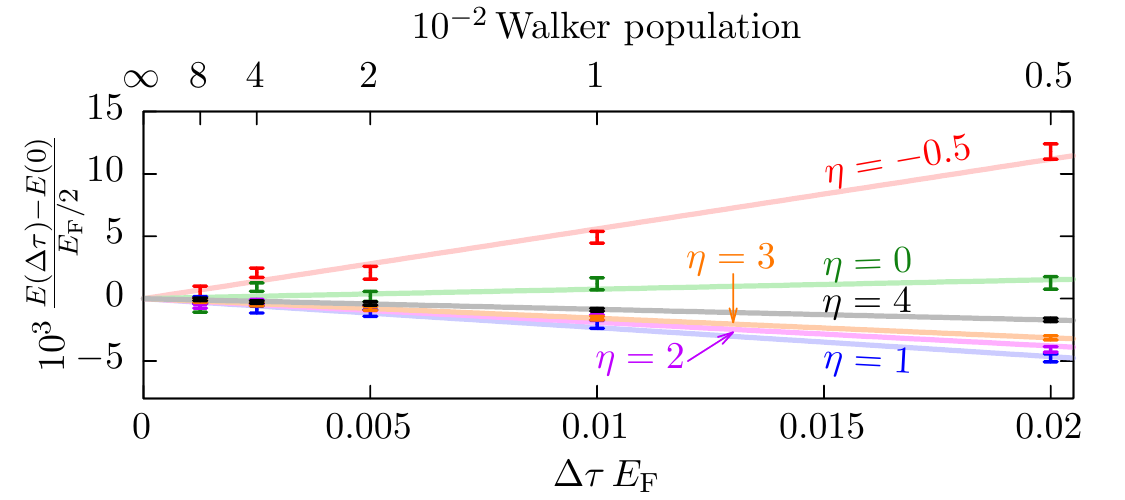}
\caption{(Color online) Variation of the dimensionless ground-state
  energy per particle with DMC time step and walker population with interaction 
  parameter $\eta=\ln(\kF a)$ in the zero-range limit $\kF^2 \Reff^2=0$. The 
  straight lines show a weighted least squares fit to the data points.}
\label{fig:timestep-2d}
\end{figure}

We simultaneously extrapolate to zero time step and infinite walker
population to eliminate the bias of the DMC algorithm, resulting from
the application of the imaginary time evolution operator
$\ee^{-\hat{H} \Delta\tau}$ at finite time steps $\Delta \tau$ on a
trial wave function represented by a finite number of walkers
\cite{Foulkes2001,Lee2011}. The absolute value of the gradient of the
energy with respect to time step is expected to be proportional to the
local energy variance, which measures the error made in using the
trial wave function to describe the true ground state
\cite{Lloyd-Williams2015,Whitehead2016}. We simultaneously reduce the
time step by a factor of 2 and increase the walker population by the
same factor, as seen in \figref{fig:timestep-2d}, where we demonstrate our
procedure for several values of the interaction parameter $\eta =
\ln(\kF a)$ for $\kF^2 \Reff^2 = 0$. As expected, the variation with
time step is smallest for $\eta = 4$, where the trial wave function
accurately captures the weakly interacting ground state of the system,
and gradually increases as $\eta$ is decreased. The gradient changes
sign near $\eta = 0$ before attaining its maximum magnitude for the
interactions parameters studied at $\eta=-0.5$. The error introduced
in the dimensionless ground-state energy by using a finite time step
and walker population increases from $0.5 \mytimes 10^{-1} \Delta \tau
\EF$ for $\eta=4$ to $2.5 \mytimes 10^{-1} \Delta \tau \EF$ for
$\eta=-0.5$. Extrapolation is therefore essential for time steps
$\Delta \tau \EF > 4 \mytimes 10^{-3}$ where this exceeds the
uncertainty in the extrapolation $< 10^{-3}$.

\subsection{System size extrapolation}
\label{finitesize-2d}

\begin{figure}
  \includegraphics[width=\linewidth]{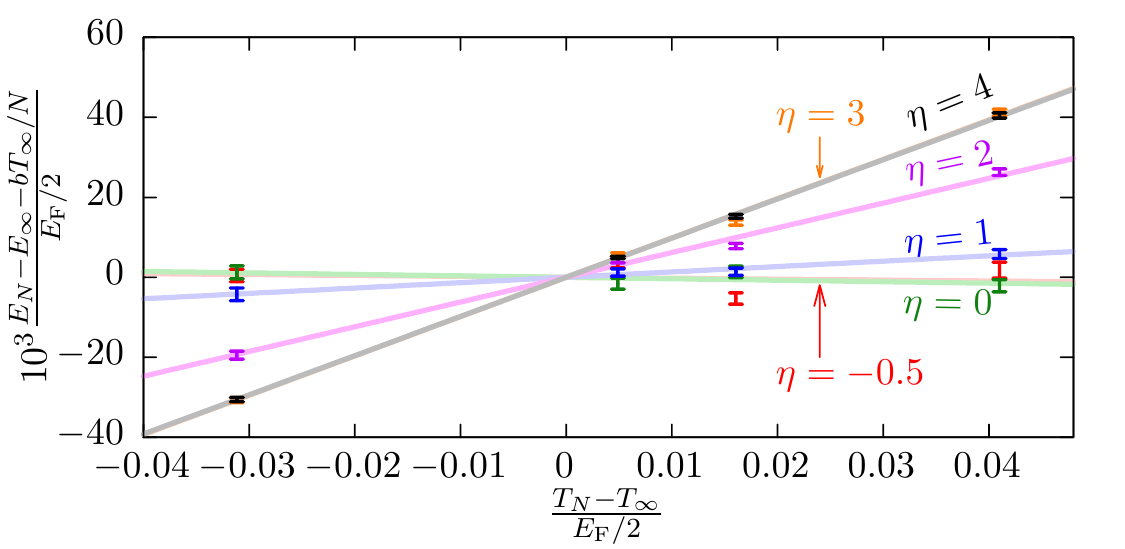}
  \includegraphics[width=\linewidth]{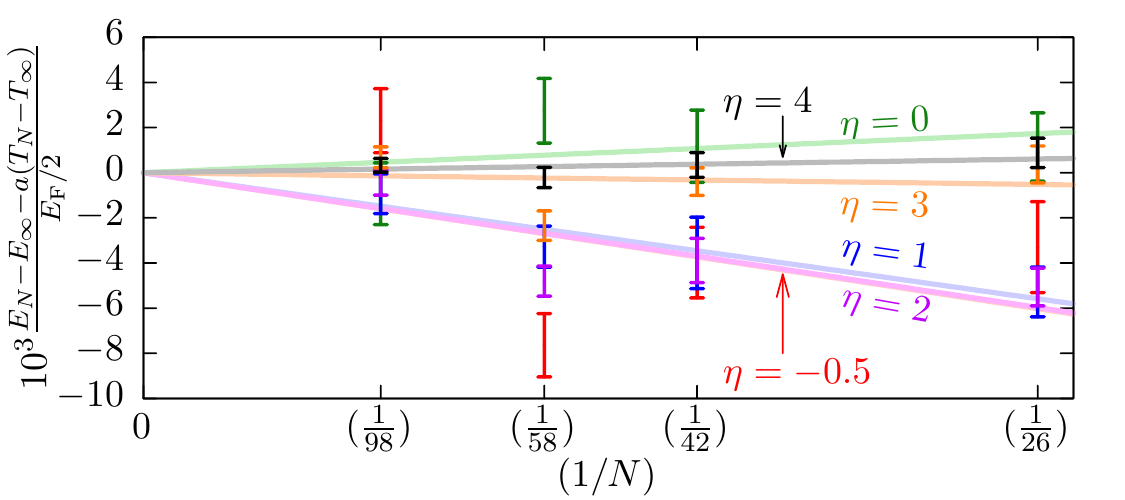}
\caption{(Color online) Variation of the dimensionless ground-state energy per 
  particle with the difference in the noninteracting kinetic energy for a 
  finite and infinite system (top) and the number of particles (bottom) for 
  various interaction parameters $\eta=\ln(\kF a)$ in the zero effective-range 
  limit $\kF^2 \Reff^2=0$. The straight lines show a weighted least squares fit 
  to the data points.}
\label{fig:finitesize-2d}
\end{figure}

Our quantum Monte Carlo algorithm simulates a finite number of
particles $N$, placed at discrete momentum vectors. This introduces a
systematic error in the kinetic energy, which in two dimensions is
proportional to $(1/N)^{\frac{5}{4}}$ \cite{Drummond2008}. 
The bias is corrected by performing simulations with varying particle number 
and extrapolating $N \rightarrow \infty$ by fitting the energy per particle 
of the $N$-particle system, $E_N$, to the formula \cite{Foulkes2001},
$$E_N = E_\infty + a (T_N - T_\infty) + b T_\infty/N,$$
where $E_\infty$ is the energy per particle for an infinite system, 
$T_N$ is the kinetic energy of a non-interacting $N$-particle system, the 
coefficient $a$ captures the rapid oscillations from the discreteness of the 
wave vectors, while $b$ captures residual finite size effects that are expected 
to fade away as $N \rightarrow \infty$, and $T_\infty = \frac{1}{2} \EF$.

We perform simulations with $N=\{26,42,58,98\}$ particles, and 
the procedure is illustrated for $\kF^2 \Reff^2 =0$ and several values of 
the interaction parameter $\eta = \ln(\kF a)$ in \figref{fig:finitesize-2d}. 
With simulations at four different number of particles, the predicted error in 
the extrapolate includes both statistical and systematic contributions. 
As shown in the top graph, oscillations are large in the weakly interacting BCS
regime, corresponding to large values of $\eta$, where the 
ground-state wave function is close to that of the noninteracting
system and the variation in the dimensionless ground-state energy is
up to $0.09$ over the range of particle numbers studied. As the value
of $\eta$ is reduced, the interparticle interaction becomes stronger
and we approach the BEC regime consisting of tightly bound fermion
pairs, washing out the finite size effects for $\eta \leq 0$.

As demonstrated in the bottom graph, the residual finite-size effects are
an order of magnitude smaller with a maximum variation in the
dimensionless ground-state energy of only $6 \mytimes 10^{-3}$. We
conclude that finite size extrapolations are essential to achieve a
target accuracy for the dimensionless ground-state energy of
$10^{-3}$, particularly in the BCS regime where oscillations in the
kinetic energy lead to finite-size errors up to $5 \mytimes 10^{-2}$.

\end{appendix}

\bibliographystyle{apsrev4-1}
\bibliography{library}

\end{document}